\documentclass[aps,prb,twocolumn,groupedaddress]{revtex4}%
\usepackage{amsmath}
\usepackage{graphics}
\usepackage{graphicx}
\usepackage{epsfig}
\usepackage{amssymb}
\usepackage{amsmath}
\usepackage{amsfonts}
\providecommand{\U}[1]{\protect\rule{.1in}{.1in}}

\begin{document}
\title{Cooper-pair size and binding energy for unconventional superconducting systems}
\author{F. \surname{Din\'{o}la Neto}$^{1}$}
\email{dinollaneto@gmail.com}
\author{Minos A. \surname{Neto}$^2$}
\author{Octavio D. Rodriguez \surname{Salmon}$^2$}
\affiliation{$^1$Centro
Universit\'{a}rio do Norte - UNINORTE\\Rua Huascar de Figueiredo, 290, Centro\\
69020-220, Manaus, AM, Brazil.\\$^2$Universidade Federal do Amazonas
- UFAM, \\ Av. General Rodrigo Oct\'{a}vio Jord\~{a}o Ramos 3000 -
Coroado I, 69077-000, Manaus, AM, Brazil.}

\begin{abstract}

The main proposal of this paper is to analyze the size of the Cooper
pairs composed by unbalanced mass fermions from different electronic
bands along the BCS-BEC crossover and study the binding energy of
the pairs. We are considering an interaction between fermions with
different masses leading to an inter-band pairing. In addiction to
the attractive interaction we have an hybridization term to couple
both bands, which in general acts unfavorable for the pairing
between the electrons. We get first order phase transitions as the
hybridization break the Cooper pairs for the the $s$-wave symmetry
of the gap amplitude. The results show the dependence of the
Cooper-pair size as a function of the hybridization for $T=0$. We
also propose the structure of the binding energy of the inter-band
system as a function of the two-bands quasi-particle energies.

\end{abstract}
\maketitle

\section{Introduction}

The unconventional superconducting systems remain an open subject in
the condensed matter physics. The pairing mechanism by which
electrons are bound together to form pairs in these systems is one
of the main issues. The superconducting (SC) state arises from
electron pairing and the onset of long-range phase coherence where
the phenomena may occur at different temperatures or other tuning
parameters. For conventional SC materials the pairing arises
indirectly by the exchange of phonons of the crystal lattice and
shows a Cooper pair size around $\xi_p\sim$10$^3-10^4${\AA}. For the
unconventional SC materials the superconducting pairing mechanism is
not completely established. Their short Cooper pair size in
comparison with conventional materials ($\xi_p\sim10^0-10^1${\AA})
have influenced theoreticians to introduce new hypotheses to explain
the electron pairing via, for example, spin fluctuations and
short-range spin waves \cite{monthoux, pines, monthoux2}.

The $s$-wave single-band models are able to describe pure metallic
superconductors, alloys and systems of cold atoms\cite{regal}, where
it is possible to study the superfluid ground state in all ranges of
attractive interactions from the BCS scenario (weak coupling) to the
Bose-Einstein Condensate (strong coupling). In the cold atoms case
it is possible to observe in the strong coupling limit the
occurrence of Bose-Einstein Condensate (BEC) that is obtained by the
single-band model. However to study more complex systems like the
heavy fermion metals, the single-band model is not enough to explain
all the experimental features observed. The heavy fermion
systems\cite{hewson} have their peculiar properties related to the
effects that the electrons show due to different electronic band
widths. Thus, multi-band models are adopted to describe these
compounds allowing the study of heavy-fermion superconductivity.
These multi-band models are similar to that which was introduced by
Suhl \textit{et al.}\cite{SMW}, with the presence of a hybridization
term between the bands, that can be tuned experimentally for example
by external pressure.

For our case, we are interested to analyze the superconductivity for
a system where the Cooper pairs are composed by fermions with
different masses. For this purpose we introduce a two-band
Hamiltonian where one band is related to a free electrons and the
second one is composed by fermions with effective mass higher than
the free electrons one. The two-band model Hamiltonian with
hybridization is given by

\begin{align} \label{1}
\mathcal{{H}} =\!\underset{{k,\sigma}}{\sum}(\varepsilon^{a}
_{k}a_{k,\sigma}^{\dag}a_{k,\sigma}\!+\varepsilon_{k}^{b}
b_{k,\sigma}^{\dag}b_{k,\sigma})\!+\!V\underset{k\sigma}{\sum}\left(
a_{k\sigma}^{\dag}b_{k\sigma
}\!+\!b_{k\sigma}^{\dag}a_{k\sigma}\right)\nonumber\\
\!-U\underset{kk'}{\sum
}a_{k\uparrow}^{\dag}b_{-k\downarrow}^{\dag}b_{-k'\downarrow}a_{k'\uparrow},
\end{align}
where $a_{k\sigma}^{\dagger}$ and $b_{k\sigma}^{\dagger}$ are
creation operators for the light $a$ and the heavy $b$
quasi-particles, respectively, with two hyperfine states labeled as
$\sigma=\uparrow$ (up spin) or $\downarrow$ (down spin). The
energies $\varepsilon^{a,b}_{k}={\sum
}_{j}t_{ij}^{a,b}e^{ik(r_{i}-r_{j})}$ are the dispersion relation of
each band with $t_{ij}^{a,b}$ being the hopping amplitudes. $V$ is
the hybridization term and $U$ is a local attractive potential among
fermions from different bands. In general, the hybridization is used
as a control parameter to explore the phase diagram, driving the
system for the superconductor quantum critical points
(SQCP)\cite{sqcp} separating normal and superconducting phases.

\section{The Green's Function of system}

We have done the Hartree-Fock mean field approximation to decouple
higher order Green's functions generated by the equations of motion
yielding a closed system of equations. The relevant wave-vector and
frequency dependent Green's functions are given by

\begin{eqnarray}
\label{2}\langle\langle{a_{k\sigma}};a_{k\sigma}^{\dagger}\rangle\rangle
= {\frac{W_{k,b}^{2-}\Omega_{k,a}^+-\Delta_{ab}^{2}
\Omega_{k,b}^+-V^{2}\Omega_{k,b}^-}{2\pi P(\omega)}},
\end{eqnarray}

\begin{align}
\label{3}\langle\langle{b_{k\sigma}};b_{k\sigma}^{\dagger}\rangle\rangle=
{\frac{W_{k,a}^{2-}\Omega_{k,b}^+-\Delta_{ab}^{2}%
\Omega_{k,a}^+-V^{2}\Omega_{k,a}^-}{2\pi P(\omega)}},
\end{align}
and
\begin{align}
\label{4}\langle\langle{b_{-k-\sigma}^{\dagger}};a_{k\sigma}^{\dagger}\
\rangle\rangle =\!{\frac{\!\Delta_{ab}(\Delta_{ab}^2\!-\!V^2 \!-\!
\Omega_{k,a}^+\Omega_{k,b}^-)}{2\pi P(\omega)}},
\end{align}
where $\Omega_{k,(a,b)}^\pm=(\omega \pm \varepsilon^{a,b}_{k})$,
$W_{k,(a,b)}^{2\pm}=(\omega^2 \pm \varepsilon^{2 a,b}_{k})$ and
$P(\omega)$ the polynomial
\begin{align}\label{6}
P(\omega)=\omega^{4}-\left[
{\varepsilon^{a}_{k}}^{2}+{\varepsilon^{b}_{k}}^2+2(\Delta _{ab}^{2}+V^{2})\right]\omega^{2}\nonumber\\
+[\varepsilon^{a}_{k}\varepsilon^{b}_{k}-(V^{2}-\Delta_{ab}^{2})]^2.
\end{align}

For this scenario $\Delta_{ab}$ is the inter-band supeconductor
order parameter, given by:
\begin{align}\label{8}
\Delta_{ab}%
=U\sum_{k}\langle a_{k\sigma}^{\dag}b_{-k-\sigma}^{\dag}\rangle.
\end{align}
We insert a simple notation to represent the inter-band pairing
correlation functions as
\begin{align}
\label{10} \phi_{ab}(k)=\langle
a_{k\sigma}^{\dag}b_{-k-\sigma}^{\dag}\rangle.
\end{align}
The anomalous Greens's function calculated
$\langle\langle{b_{-k-\sigma}^{\dagger}};a_{k\sigma}^{\dagger}\rangle\rangle$
allow one to obtain the correlation function previously described
$\phi_{ab}(k)$ through the Fluctuation-Dissipation
Theorem\cite{tyablikov} as follows
\begin{align}\label{12}
\phi_{ab}(k)=\int d\omega f_{FD}(\omega) \Im
m\langle\langle{b_{-k-\sigma}^{\dagger}};a_{k\sigma}^{\dagger}\rangle\rangle,
\end{align}
where $f_{FD}(\omega)=[\exp(\beta\omega)+1]^{-1}$ is the Fermi-Dirac
function ($\beta=1/k_BT$ where $k_B$ is the Boltzmann constant and
$T$ the temperature). After perform these calculations we get (for
$T=0$):

\begin{align}\label{14}
\phi_{ab}(k)=\frac{\Delta_{ab}}{2(\omega_{1k}^2\!-\!
\omega_{2k}^2)}\underset{{i=1}}{\overset{2}\sum}(\!-1\!)^{i\!-\!1}
\frac{{(\omega_{ik}^{2}\!-\varepsilon^{a}_{k}\varepsilon^{b}_{k}\!-\Delta_{ab}^{2}+V^2})}{\omega_{ik}}.
\end{align}
The roots of the polynomial $P(\omega)$ in Eq.(\ref{6}) yield the
poles of the Green's functions and determine the energy of the
excitations of the two-band superconductor. These are given by,
\begin{equation} \label{15}
\omega_{1,2k}=\sqrt{A(k)\pm\sqrt{B(k)}},
\end{equation}
with
\begin{equation}
A(k)=\frac{{\varepsilon^{a}_{k}}^{2}+{\varepsilon^{b}_{k}}^2+2(\Delta_{ab}^{2}+V^{2}%
)}{2}\nonumber
\end{equation}
and
\begin{align}
B(k)=\left[\frac{({\varepsilon^{a}_{k}}^{2}-\varepsilon^{b 2}_{k})}%
{2}\right]^{2}+V^{2}\left[(\varepsilon^{a}_{k}+\varepsilon^{b}_{k})^{2}\right]\nonumber \\ +\Delta_{ab}^{2}\left[(\varepsilon^{a}_{k}-{\varepsilon^{b}_{k})}^{2}%
+4V^{2}\right].\nonumber
\end{align}

In our analysis, we assume that the bands display the same shape
being homotetic, i.e., $\varepsilon^{b}
_{k}=\alpha\varepsilon^{a}_{k}$ with
$\varepsilon^{a}_{k}=k^{2}/2m_a-\mu$, where $\alpha$ is the ratio of
the effective masses of the quasi-particles in the two bands given
by $\alpha=m_a/m_b$, with $m_a$ and $m_b$ being the masses of the
particles of the wide and the narrow band, respectively. The
chemical potential $\mu$ is the same for both bands. Since the
$b$-electrons from the narrow band are heavier, $\alpha<1$. We
consider the total number of electrons $N\!=\!\sum_{k}(\langle
n^{a}_k\rangle+\langle n^{b}_k\rangle)$ as fixed, where $\langle
n^{a}_k\rangle=\langle a_{k}^{\dag}a_{k}\rangle$ and $\langle
n^{b}_k\rangle=\langle b_{k}^{\dag}b_{k}\rangle$ are calculated
similarly as done in Eq. (\ref{12}) for the anomalous Green's
functions.

We now obtain the number and the gap equations for a three
dimensional system with $s$-wave order parameter symmetry. These
must be solved self-consistently. The number equation for $T=0$ is
obtained from the propagators (\ref{2}) and (\ref{3}). It is given
by,

\begin{align}\label{16} & N=\frac{k_{F}^{3}}{4\pi^{2}}
\int_{-\overline{\mu}}^{\infty} \sqrt {(x+\overline{\mu})} \left\{
1-\frac{x}{2(\overline{\omega}_{1x}^{2}-\overline{\omega}_{2x}^{2})}
\times\right. \nonumber\\ &  \left.
\underset{{i=1}}{\overset{2}\sum}(\!-1\!)^{i\!-\!1}
\frac{(\overline{\omega}_{ix}^{2}-\overline{\Delta}_{ab}^2+
\overline{V}^{2}-\alpha x^{2})(1+\alpha)}{\overline{\omega}_{ix}}
\right\} dx, \end{align} where $\omega_{1,2x}$ are the excitation
energies of the system given by (\ref{15}). The sum over $k$ was
changed into an integral, and we introduced  a dimensionless
variable $x$. The over-bar in a given quantity means that it is
normalized by the Fermi energy $E_{F}$.

In a pure inter-band scenario i.e., without any contribution of a
intra-band attractive interaction, it is observed a first-order
phase transition at low temperatures and a second-order phase
transition for higher temperatures\cite{PLA}. In consequence, a
tricritical point (TCP) in the phase diagram was found similar to
that observed experimentally in superconducting systems with an
applied magnetic field. However in our description we have no
presence of a magnetic field, instead we have a control parameter,
the hybridization, that can be tuned by external pressure or doping.
In general these observations indicate the possibility of measuring
discontinuities in the SC gap amplitude by applying pressure to the
system.

In the Suhl model the hybridization between the bands was neglected
but they included an ``inter-band" term that creates(annihilates) an
intra-band pair from one band and annihilates(creates) an intra-band
pair from the other one. Thus the model that we adopted in this
section is different because we do not consider any kind of
intra-band pairs. Our attractive inter-band interaction creates a
pair composed by quasi-particles from different species as pointed
out by V. Liu and F. Wilczek\cite{liu} to propose a new state of
matter in which the pairing interactions create a gap within the
interior of a large Fermi ball, while the exterior surface remains
gapless. In general, the pairing occurs between two species whose
Fermi surfaces do not match simply because their densities or
effective masses differ. This possibility arises in solids for
example if the electron populations in the two bands are unbalanced.

The hybridization mechanism is responsible for the mixing among the
quasi-particles from different bands. In metallic systems, such as
transition metals\cite{kishore}, inter-metallic compounds and heavy
fermions\cite{jullien}, this term arises from the mixing of the
wave-functions of the quasi-particles in different orbitals through
the crystalline potential. For another systems as the problem of
color superconductivity, it is the weak interaction that allows the
transformation between up and down-quarks\cite{sedrakian,huang}. For
a system of cold fermionic atoms in an optical lattice with two
atomic states, the hybridization is related to Raman transitions
with an effective Rabi frequency which is linearly proportional to
$V$ \cite{liu}. Then, the physical origin of the $V$-term is
different for each case. Nevertheless, the main point is that at
least in inter-metallic systems, the hybridization can be easily
controlled by pressure or doping\cite{leticie} allowing one to
explore their phase diagrams. That is one of the most interesting
aspects that lead us to use the these quantities as external control
parameters, and that can be used to find for some systems the
superconducting quantum critical point (SCQP) related to a
second-order phase transitions.

From the Eqs. (\ref{8}) and (\ref{14}) we write the inter-band gap
equation for $T=0$ as
\begin{align}\label{20}
\frac{1}{U}\!=\!\underset{k}\sum\frac{1}{2(\omega_{1k}^2\!-\!
\omega_{2k}^2)}\underset{{i=1}}{\overset{2}\sum}(\!-1\!)^{i\!-\!1}
\frac{{(\omega_{ik}^{2}\!-\varepsilon^{a}_{k}\varepsilon^{b}_{k}\!-\Delta_{ab}^{2}+V^2})}{\omega_{ik}}.
\end{align}

We are interested to solve the gap equation in (\ref{20}) and
calculate the inter-band Cooper-pair size and to analyze the effects
of $V$ on these pairs along all spectra of interactions. Therefor we
perform the renormalization technique in the gap equation in Eq.
(\ref{20}) to remove the divergence of the integral in the strong
coupling limit as made in Ref.\cite{randeria,samelos}. The standard
solution of this problem is to transfer this divergence to the
$s$-wave scattering length $a_{s}$ \cite{samelos}. This quantity can
be positive (strong coupling) or negative (weak coupling), and
diverges at the unitarity limit ($1/a_{s}=0$) as the systems crosses
over BCS to the BEC limit. The scattering length parameter allows
one to describe all regimes of attractive interactions between the
quasi-particles. Thus, after performing the renormalization
procedure in Eq.(\ref{20}) we obtain

\begin{align}\label{21}
\frac{\textmd{-1}}{k_{F}a_{s}}\!=\!\frac{1}{\pi}\!\int_{-\overline{\mu}}^{\infty}
\frac{\sqrt{(x\!+\!\overline{\mu})}}{(\overline{\omega}
_{1x}^{2}\textmd{-}\overline{\omega}_{2x}^{2})}\left[\underset{{i=1}}{\overset{2}\sum}(\textmd{-1}\!)^{i\textmd{-}1}
\frac{(\overline{\omega}_{ix}^{2}\textmd{-}\alpha
x^{2}\textmd{-}\overline{\Delta}_{ab}^{2}\!+\!\overline{V}^{2})}{\overline{\omega}_{ix}}\right.
\nonumber\\
\left. \textmd{-}\frac{2}{(1+\alpha)\sqrt{(x+\overline{\mu})}}
\right] dx,
\end{align}
where in this representation, $1/k_{F}a_{s}$ plays the role of a
dimensionless coupling constant, and $k_{F}=\sqrt{2mE_{F}}$ is the
Fermi wave vector. Notice that the integral limits in Eq. (\ref{21})
are [$-\overline{\mu},\infty$]. Similarly to the intra-band case,
when $1/k_{F}a_{s} \rightarrow-\infty$ one obtains the weak coupling
regime (BCS limit), while for $1/k_{F}a_{s} \rightarrow+\infty$ the
system reaches the strong coupling limit (BEC limit).

The inter-band pair size is obtained using the definition
\cite{randeria} based on the $\langle
r^2\rangle={(\xi_p^{inter})}^{2}$ calculation

\begin{align}\label{22}
{(\xi_p^{inter})}^{2}=\frac{\int d^3k
\;\phi_{ab}^*(k)\nabla_k^2\phi_{ab}(k)}{\int d^3k \;\phi_{ab}^*(k)
\phi_{ab}(k)},
\end{align}
where the inter-band pair wave function $\phi_{ab}(k)$ is given by
the Eq. (\ref{14}).

\section{Results and discussions}

Now, the integral that results from Eq. (\ref{22}) is solved
together with the number equation (\ref{14}) and the renormalized
gap equation (\ref{21}) self-consistently.

\begin{figure}[t] \centering
\includegraphics[angle=0,scale=1.0,height=7.5cm]{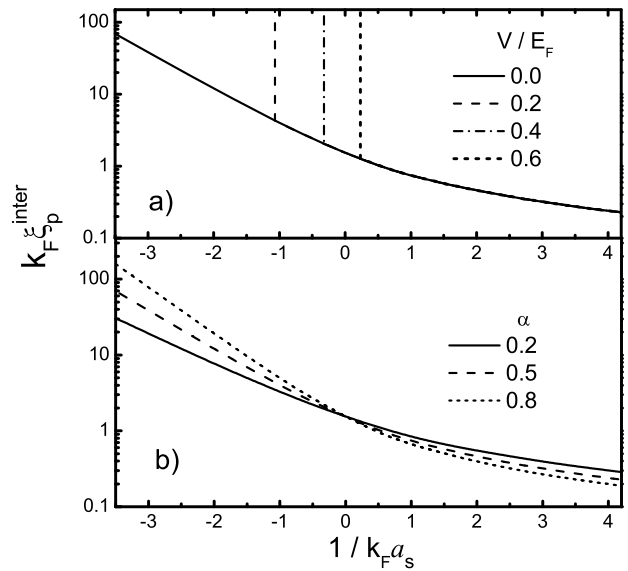}
\caption{a) The inter-band Cooper-pair size dependence of attractive
interaction ($1/k_Fa_s$) for several values of hybridization
$V/E_F$. b) The inter-band Cooper-pair size dependence of attractive
interaction ($1/k_Fa_s$) for several values of mass ratio $\alpha$.
}\label{fig1}
\end{figure}

\begin{figure}[t] \centering
\includegraphics[angle=0,scale=1.0,height=7.5cm]{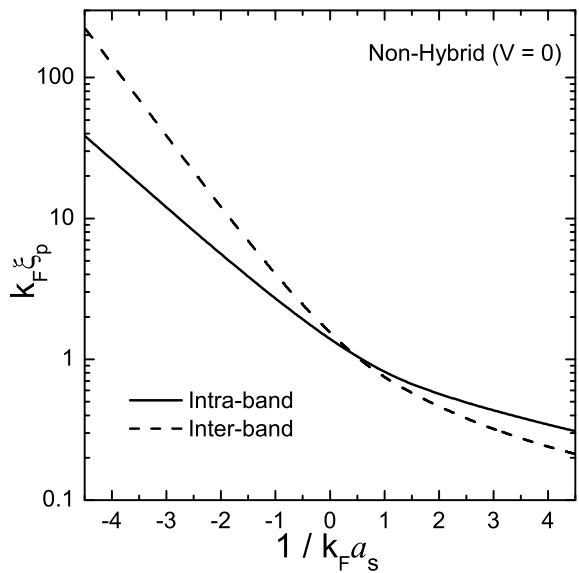}
\caption{The intra and inter-band Cooper-pair size dependence of
attractive interaction ($1/k_Fa_s$) for the non-hybrid case
$V/E_F=0$ and $\alpha=0.5$.}\label{fig2}
\end{figure}

In Fig. \ref{fig1} a) (upper graph) we plot the numerical result for
$k_F\xi_p^{inter}$ for several values of hybridization $V$. We
observe that for non-hybrid case the numerical solution for
$\xi_p^{inter} \times (1/k_Fa_s)$ is quite similar as that obtained
the intra-band case \cite{dinola3}, being a monotonically decreasing
function of the attraction and proportional to $1/\Delta_{ab}$,
going from the exponentially large value in the BCS limit. The
numerical data for $\overline{\Delta}_{ab}$ and $\overline{\mu}$
were obtained from self-consistently solutions of integrals in Eq.
(\ref{16}) and Eq. (\ref{21}). For hybrid case ($V>0$) we observe
that the curves are quite different from the intra-band scenario:
For a fixed value of hybridization we must achieve a minimum
strength in the attractive interaction to get a solution for the
problem. For example, for $V=0.2$ does not exist solutions for
$1/k_Fa_s < -1$, i.e., for a fixed attractive interaction there is a
characteristic value of hybridization which leads the system from a
SC state to the normal one. This fact is related to the First-order
transitions that occur in the inter-band scenario as observed in
Ref. \cite{PLA}. From another point of view the attractive
interaction must overcome the quantum barrier imposed by the
hybridization, i.e. the superconducting state arises when
${\Delta}_{ab}>V$. The First-order transition in the BCS-BEC
crossover for the inter-band case was firstly pointed out in
Ref.\cite{dinola}. Nevertheless the convergence observed only in the
BEC limit for the intra-band case is now observed also in the BCS
and BEC limits for inter-band case. In the BEC limit we observe a
convergence of all solutions for different hybridization strength
going to the same size of the two-body bound state observed for
$V=0$.

In the Fig. \ref{fig1} b) (lower graph) we plot $k_F \xi_p^{inter}$
for several values of mass ratio $\alpha$. In the BCS regime, for
systems where $\alpha$ is close to 1, as our result for
$\alpha=0.8$, the Cooper pair is composed by fermions with balanced
masses as those found in conventional systems. The size of these
pairs are much bigger than those formed by unbalanced masses as our
result for $\alpha=0.2$. The short size of the Cooper-pairs for the
case where $\alpha$ is small, it is comparable to the heavy fermions
superconductors (HFSC) and the HTSC where $\alpha \ll 1$ and is in
agreement with experimental observations. For conventional BCS SC
materials the Cooper pair size is around
$\xi_p\sim$10$^3-10^4${\AA}, while for the unconventional SC
materials the Cooper pair size is around $\xi_p\sim10^0-10^1${\AA}.

At this point it is possible to perform a comparison between
intra-band \cite{dinola3} and inter-band systems. In Fig. \ref{fig2}
we plot the results for $\xi_p^{intra}$ and $\xi_p^{inter}$ for
non-hybrid $V=0$ case. We observe that in the BCS limit the
inter-band Cooper-pair is much bigger than the intra-band one. This
fact corroborates with another theoretical observations
\cite{dinola, igor}. This fact indicates that for a same attractive
interaction strength, the intra-band gap amplitude $\Delta$ is
bigger than $\Delta_{ab}$. In the BCS-BEC crossover region
($-1<1/k_Fa_s<+1$) we observe that close to the unitarity
($1/k_Fa_s=0$) both sizes coincide $\xi_p^{intra}=\xi_p^{inter}$. In
this region of attractive interactions a Cooper-pair composed of
fermions with same mass has the same size of the other one composed
of an unbalanced massive fermion system and $\Delta \sim
\Delta_{ab}$. This observation indicates that in the BCS-BEC
crossover the $\alpha$ factor (masses ratio) is not relevant to
determine the Cooper-pair size for both kind of interaction (intra
or inter-band). At this point for higher interactions there occurs a
significant change in the major size of Cooper pairs. Now when the
system is driven to the BEC limit, the inter-band Cooper-pair size
becomes slightly smaller than the intra-band one, i.e., $\Delta <
\Delta_{ab}$. However this fact does not change the main feature of
the system: the first-order phase transition of inter-band scenario.

\begin{figure}[t] \centering
\includegraphics[angle=0,scale=1.0,height=7.5cm]{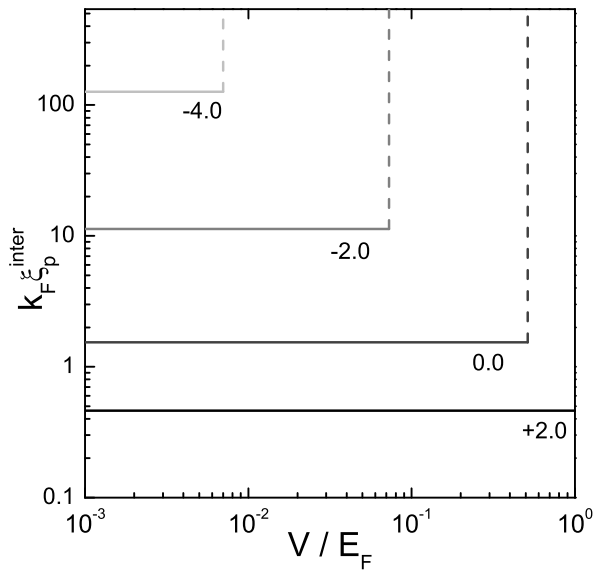}
\caption{The inter-band Cooper-pair size dependence of hybridization
$V/E_F$ for several values of attractive interaction $1/k_Fa_s$ for
BCS ($1/k_Fa_s<0$) and BEC limits ($1/k_Fa_s>0$). For this scenario
discontinuities are observed.}\label{fig3}
\end{figure}

In the Fig. \ref{fig3} we analyze with more attention the inter-band
Cooper-pair size dependence with hybridization $V/E_F$ for several
values of attractive interaction for BCS ($1/k_Fa_s<0$) and BEC
limits ($1/k_Fa_s>0$). In this case the system behaves similarly as
the for finite temperatures: generally the Cooper-pair size is
independent of temperature for conventional superconductors
\cite{mourachkine}. Our results show that $\xi_p^{inter}$ is not
modified by hybridization until the characteristic $V$. When $V$
reaches the condition $V\geq\Delta_{ab}$ the inter-band Cooper pair
is abruptly broken and is observed one of the most relevant features
of inter-band superconductivity, the first-order phase transition.
However the order of magnitude of $k_F\xi_p^{inter}$ for both BCS
and BEC limits is not altered in comparison to intra-band case: In
the BCS limit where $1/k_Fa_s<0$, $k_F\xi_p^{inter}\sim10^1-10^2$.
In BEC limit we observe that even for decades of order of magnitude
for $V$, $\xi_p^{intra}$ keeps unaltered around
$k_F\xi_p^{inter}\sim10^{\textmd{-}1}-10^{0}$.

\subsection{Spectral analysis and the binding energy of system}

\begin{figure}[t] \centering
\includegraphics[angle=0,scale=1.0,height=8cm]{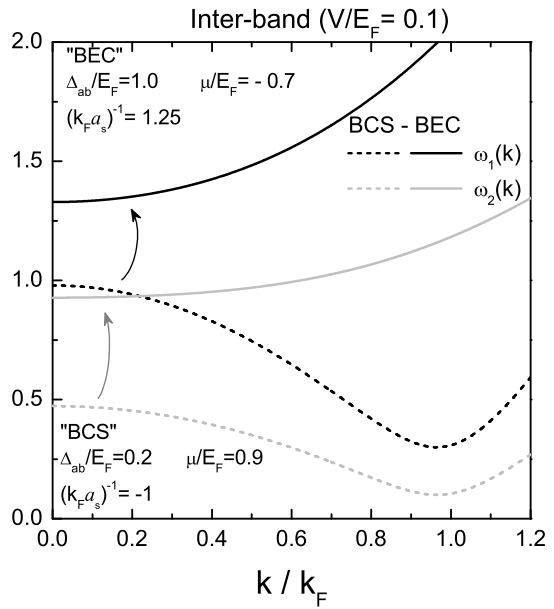}
\caption{Dispersion relation of the inter-band excitations
$\omega_{1}(k)$ and $\omega_{2}(k)$ in the BCS weak coupling regime
and strong coupling BEC regime for $V/E_F=0.1$ and
$\alpha=0.5$.}\label{fig4}
\end{figure}

In this section we study the spectra of the system. From numerical
data we can perform a spectral analysis of the quasi-particles
$\overline{\omega}_{1}(k)$ and $\overline{\omega}_{2}(k)$ along the
BCS-BEC crossover at $T=0$. For the inter-band case for a fixed
hybridization $V/E_F=0.1$ we plot the results in Fig. \ref{fig4}. In
the BCS limit (dotted lines) the typical dip close to the Fermi wave
vector for $\overline{\omega}_1(k)$ and $\overline{\omega}_2(k)$ is
observed. In this picture $\Delta_{ab}$ reaches the minimal value to
give rise the inter-band SC state. In the strong coupling case (full
lines) shown in Fig. \ref{fig4} the dispersion relations are similar
to those observed in the intra-band case: The effective Bose
particles with a quadratic dispersion. In this case $\Delta_{ab}>V$
and there are no gapless or dip point in the spectra. Similarly to
intra-band case, the gap at $k = 0$ can be associated to the
dissociation energy of the effective bosons formed now by the
strongly coupled pairs of fermionic quasi-particles with different
masses.

\begin{figure}[t] \centering
\includegraphics[angle=0,scale=1.0,height=8cm]{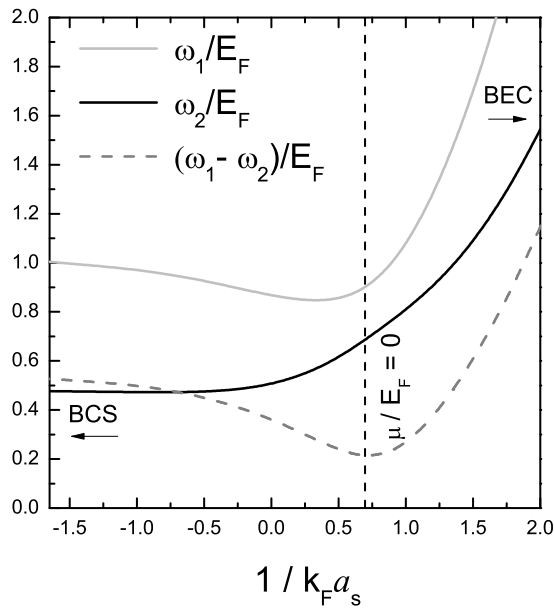}
\caption{The energies $\overline{\omega}_{1}$ and
$\overline{\omega}_{2}$ for $k=0$ as a function of the attractive
interaction in the BCS weak coupling regime and strong coupling BEC
regime and the difference $(\omega_{1}-\omega_{2})/E_F$ for
$V/E_F=0.1$ and $\alpha=0.5$.}\label{fig5}
\end{figure}

In the Fig. \ref{fig5} we plot the energies $\overline{\omega}_{1}$
and $\overline{\omega}_{2}$ for $k=0$ as a function of the
attractive interaction. We are interested in observing the
dissociation energy of the effective bosons of the system as a
function of $1/k_Fa_s$. For this inter-band scenario, in the weak
coupling limit $\overline{\omega}_{1}=1$, i.e., ${\omega}_{1}=E_F$
and $\overline{\omega}_{2}=0.5$ as expected for the BCS limit
considering $\alpha=0.5$. Still in the BCS limit the chemical
potential is around the Fermi energy of the system $\mu \sim E_F$.
However, as the attractive potential increases, $\mu$ must be
adjusted to keep fixed the total particle number in Eq.(\ref{16}).
As the chemical potential get closer to $\mu=0$ the system composed
initially by fermions is driven to a system where these fermions are
tightly bound and superconducting ground state is composed now by
effective bosons which behaves like as free particles with a
quadratic dispersion as shown in Fig. \ref{fig4}. Nevertheless,
differently of the intra-band case where the binding energy was
purely the $\overline{\omega}_{2}$ that displays a dip for $\mu=0$,
for the inter-band case we found that the binding energy $E_b$ of
the system is written as the difference between both branches of
energies, i.e, $E_b=\omega_{1}- \omega_{2}$, that dip and reaches
the minimum value when $\mu=0$ (dotted vertical line) as shown in
the Fig. \ref{fig5}. Notice that $E_b$ does not display a dependence
with the hybridization based on our result shown in Fig. \ref{fig3}.
That is a remarkable feature and it is a significant difference in
comparison with the intra-band system where $E_b$ has a power law in
function of $V$. However $E_b$ shows a dependency with the mass
ratio $\alpha$. A clear evidence is shown the Fig. \ref{1}, where
for a fixed attractive interaction, the $\xi_p^{inter}$ get higher
as $\alpha$ increases. Performing an expansion of $E_b$ as a
function of $\alpha$ we get
\begin{eqnarray}\label{23}
E_b(\alpha, \Delta_{ab}, \mu)=A_0- A_1\alpha+\mathcal{O}(\alpha^2),
\end{eqnarray}
where
\begin{equation}
A_0(\Delta_{ab}, \mu)=\frac{A^{+}(\Delta_{ab},
\mu)-A^{-}(\Delta_{ab},
\mu)}{2}\nonumber\\
\end{equation}
and
\begin{equation}
A_1(\Delta_{ab},\mu)=\frac{2\mu\Delta_{ab}^2}{\sqrt{{\mu^2}+4\Delta_{ab}^2}}\left(\frac{1}{A^{+}(\Delta_{ab},
\mu)}-\frac{1}{A^{-}(\Delta_{ab}, \mu)}\right), \nonumber
\end{equation}
with the coefficients
\begin{equation}
A^{\pm}(\Delta_{ab},\mu)=\sqrt{2\mu^2+4\Delta_{ab}^2\pm
2\mu\sqrt{\mu^2+4\Delta_{ab}^2}}.\nonumber
\end{equation}
We may notice that the Eq. (\ref{23}) is suitable for a little
$\alpha$, where the term $\mathcal{O}(\alpha^2)$ can be neglected.
Also in the Eq. (\ref{23}) we observe that while $\alpha$ increases
$E_b$ linearly decreases.

\section{Conclusions}

In this work, we have studied the effects of hybridization $V$ on
the superconducting ground state and on the inter-band Cooper-pair
size. We adopted the two-band model with an inter-band attractive
interaction. Using the Hartree-Fock mean field approximation we
obtain the normal and anomalous Green's functions which are then
used to determine the gap and number equations, and the Cooper-pair
size equation. These equations are then solved self-consistently
separately.

In order to observe the hybridization effects on Cooper-pair size
for inter-band interactions and study all ranges of interactions
(BCS weak coupling limit and the BEC strong coupling limit) we
renormalized the gap equation, by introducing the scattering length
for the two-band problem. For $V=0$ the inter-band case shows the
expected smooth evolution for $\xi_p^{inter}$ between weak and
strong coupling limits. In the BEC limit we observe a convergence of
all solutions for different hybridization strength going to the same
size of the two-body bound state observed for $V=0$. However for a
fixed attractive interaction strength, $\xi_p^{inter}$ does not
change and keep the same size until a certain value of $V$ where
$\xi_p^{inter}$ suddenly disappears when the Cooper-pair is broken.
These facts are related to the first-order transitions observed by
the discontinuities in the inter-band gap amplitude $\Delta_{ab}$.

In the BEC limit we have observed that the difference between the
quasi-particle energies $\omega_{1}-\omega_{2}$ becomes the pair
binding energy for $k=0$. When the chemical potential goes to
$\mu=0$, it was observed the minimum in the binding energy where the
superconducting ground state is now composed by the effective
bosons. We also performed an expansion of the binding energy $E_b$
as a function of the mass ratio and was observed that while $\alpha$
increases $E_b$ decreases.

\acknowledgments The authors thank to the Brazilian Agencies, FAPEAM
and CNPq for partial financial support.

\end{document}